\documentclass[11pt]{article}
\usepackage[margin=1in]{geometry}
\usepackage{amsmath,amssymb}
\usepackage{array}
\usepackage{longtable}
\usepackage{booktabs}
\usepackage{xcolor}
\usepackage{hyperref}
\usepackage{algorithm}
\usepackage{algpseudocode}
\usepackage{graphicx}
\usepackage{authblk}
\usepackage{multirow}

\hypersetup{
    colorlinks=true,
    linkcolor=blue!60!black,
    citecolor=blue!60!black,
    urlcolor=blue!60!black
}

\title{\textbf{Algorithmic Threshold Optimization: Quantitative Modeling of Multiplier Distributions in Crash Games}}

\author{\textbf{Sourish Sarkar}}
\affil{Indian Statistical Institute}
\affil{\href{mailto:sourish.sarkar13@gmail.com}{\texttt{sourish.sarkar13@gmail.com}}}

\date{}

\begin{document}

\maketitle

\begin{abstract}
Crash is a widely played casino game that blends strategic decision-making, probability, and computational analysis, particularly from the perspective of the house. This paper presents an optimization algorithm aimed at minimizing the casino's guaranteed positive earnings while simultaneously maximizing the number of players who win some amount during a given round. Since Crash is inherently a multiplayer game, players place random investment amounts at different points in time, which makes it difficult for any individual player to judge, before the round even begins, whether their intended investment is comparatively high or low, and consequently, how much risk they are actually taking on. The core objective of the proposed algorithm is to determine an optimal stopping multiplier that balances both the casino's and the players' interests. Notably, the algorithm remains unaffected by changes in the number of players or by variation in investment amounts, although in practice casinos typically impose a fixed range on permissible investments. Beyond this, the algorithm can help professional players make more informed decisions about how much to invest and how much risk they are comfortable accepting. We further analyze the algorithm's computational complexity and its stopping-multiplier statistics to give a fuller picture of how it behaves in practice.
\end{abstract}

\section{Game Overview}

Crash is a multiplayer casino game in which a fixed cohort of players registers before each round begins. Once registration closes, the round starts at a predetermined time, and only players who registered beforehand are eligible to take part in that particular round. Each player is permitted to invest an amount that typically falls between \$1 and \$100, expressed to two decimal places, which allows for fairly fine-grained variation in stake sizes across the player pool.

At the start of a round, the multiplier is set to 1.00 and climbs continuously from there, in principle without any upper bound, again measured to two decimal places. While the round is running, any player may choose to cash out at the prevailing multiplier, an action that immediately locks in a payout proportional to that value. The round ends when the underlying engine triggers a crash at some multiplier, which we refer to throughout this paper as the \emph{stopping multiplier}. Players who have not cashed out before this happens forfeit their entire stake, while players who cash out beforehand receive a payout equal to their investment multiplied by their chosen cash-out multiplier, yielding a net profit relative to their original stake.

To illustrate, consider a player who invests \$50 and cashes out at a multiplier of 1.42, while the round's stopping multiplier later turns out to be 3.29. Since the player's cash-out multiplier is lower than the stopping multiplier, the cash-out succeeds, and the player's total return works out to:
\begin{equation}
\text{Payout} = 50 \times 1.42 = \$71
\end{equation}
for a net profit of
\begin{equation}
\text{Profit} = 71 - 50 = \$21.
\end{equation}

Some variants of the game start the multiplier at 0.00 rather than 1.00, which raises a real fairness concern: if the stopping multiplier happens to land below 1.00, every player in that round loses their entire stake regardless of strategy, since most participants would not think to cash out below unity in the first place. In these rounds, even a player who reacts instantly has little realistic chance of breaking even. This is a structural asymmetry that arguably works against the fairness the game is supposed to offer, since the outcome ends up depending almost entirely on an early, unfavorable crash rather than on anything the player actually did. This is one of the main reasons we set out to build the optimization framework described in this paper. Games built around a proportional payout pool of this kind sit closer to parimutuel wagering systems than to fixed-odds betting \cite{ottaviani2006}, and the aggregate wager and payout patterns that emerge from probability-based online gambling games of this general type have themselves been the subject of empirical study \cite{wang2019}.

\section{Algorithm}

\begin{algorithm}[H]
\caption{Adaptive Stopping-Multiplier Determination for Crash Rounds}
\label{alg:crash-stopping}
\begin{algorithmic}[1]
\Require Set of players $P = \{p_1, p_2, \dots, p_n\}$ with investments $I(p_i)$;
         minimum multiplier $m_{\min}$; maximum multiplier $m_{\max}$;
         multiplier increment $\delta$
\Ensure Stopping multiplier $M^*$, final payout sum $S$, casino payout, company profit

\State $T \gets \sum_{i=1}^{n} I(p_i)$ \Comment{Total investment pool}
\For{each player $p_i \in P$}
    \State $\pi(p_i) \gets I(p_i) / T$ \Comment{Contribution-based cash-out probability}
\EndFor
\State $R \gets P$ \Comment{Remaining (not yet cashed-out) players}
\State $c \gets m_{\min}$ \Comment{Current multiplier}
\State $S \gets 0$ \Comment{Cumulative normalized payout}
\State $M^* \gets m_{\max}$ \Comment{Default stopping multiplier if never triggered}

\While{$c \leq m_{\max}$}
    \State $K \gets \emptyset$ \Comment{Players closing at this multiplier}
    \For{each player $p_i \in R$}
        \State draw $r \sim \mathcal{U}(0,1)$
        \If{$r < \pi(p_i)$}
            \State $K \gets K \cup \{p_i\}$
        \EndIf
    \EndFor

    \If{$K \neq \emptyset$}
        \State $S \gets \min\left(S + \sum_{p_i \in K} \pi(p_i)\cdot c,\; 1\right)$
        \State $R \gets R \setminus K$ \Comment{Remove players who closed}
    \EndIf

    \State $\pi_{\max} \gets \max_{p_i \in R} \pi(p_i)$ \Comment{0 if $R = \emptyset$}
    \State $x \gets 1 - S$ \Comment{Remaining payout budget}
    \State $y \gets \pi_{\max} \cdot c$ \Comment{Worst-case next payout at current multiplier}

    \If{$y \geq x$}
        \State $M^* \gets c$
        \State \Return $M^*, S$ \Comment{Stop --- worst case would breach the cap}
    \EndIf

    \State $c \gets c + \delta$
\EndWhile

\State \Return $M^*, S$ \Comment{Loop exhausted without triggering; crash at $m_{\max}$}
\end{algorithmic}
\end{algorithm}

\section{Step-by-Step Description of the Algorithm}

At its core, the algorithm asks the same question at every multiplier tick: can the round safely continue, or has it reached the point where letting it run one more step risks paying out more than the pool can cover? Working through it step by step:

\begin{enumerate}

\item \textbf{Work out each player's share of the pool.} We first add up every player's investment to get the total pool for the round. Each player's cash-out probability is then just their own stake divided by this total, so someone who has put in a larger share of the pool is proportionally more likely to close their position at any given multiplier tick.

\item \textbf{Set up the round.} The multiplier starts at its minimum value, the running payout total starts at zero, and every player begins in the pool of players who are still active.

\item \textbf{Simulate cash-outs at the current multiplier.} For each player still in the game, we draw a random number. If it comes in below that player's cash-out probability, we treat them as having cashed out at the current multiplier.

\item \textbf{Update the running payout.} If any players cashed out this step, we add up their combined contribution (each player's probability times the current multiplier) and add it to the running total, capped at one, which represents the full pool. Those players then leave the active pool.

\item \textbf{Check the worst case.} After handling this step's cash-outs, we look at the single highest remaining cash-out probability among the players still active, then compute two numbers: how much of the payout budget is left, and what the payout would be in the worst case, if that highest-probability player cashed out right now.

\item \textbf{Decide whether to stop.} If that worst-case payout would meet or exceed what's left in the budget, the round ends immediately, and the current multiplier becomes the stopping multiplier. Letting the round continue even one more step at that point risks paying out more than the pool can sustain. If there's still comfortable room in the budget, the round keeps going.

\item \textbf{Move the multiplier forward.} If the round hasn't been stopped, the multiplier ticks up by a small fixed increment, and the process loops back to Step 3 at the new, slightly higher multiplier.

\item \textbf{Stop at the ceiling if nothing else triggers first.} If the multiplier reaches its predefined maximum without the stopping condition ever firing, the round simply ends there by default, and whatever payout total has accumulated is returned as the final result.

\end{enumerate}

The key difference from a purely reactive design is that this algorithm doesn't wait to react only after a payout cap has already been exceeded. At every step it looks ahead to the worst case first, and only lets the round continue if the pool can still absorb the largest plausible remaining payout. That gives the casino a deterministic, mathematically grounded safeguard against unbounded exposure, while still letting the round run as long as possible in the players' favor.

\section{Complexity Analysis}
\label{sec:complexity}

Let $n = |P|$ denote the total number of players in a round, and let
$k = \left\lceil \dfrac{m_{\max} - m_{\min}}{\delta} \right\rceil$
denote the total number of multiplier steps executed by the outer
\textbf{while} loop, where $m_{\min}$ and $m_{\max}$ are the minimum and
maximum multipliers and $\delta$ is the fixed multiplier increment. Following standard practice for analyzing iterative algorithms of this kind \cite{cormen2009}, we break Algorithm~\ref{alg:crash-stopping} down into its
constituent phases, summarized in Table~\ref{tab:complexity}.

\begin{table}[H]
\centering
\caption{Time and space complexity of Algorithm~\ref{alg:crash-stopping} by phase}
\label{tab:complexity}
\begin{tabular}{@{}p{4.3cm}p{3.5cm}p{2.3cm}p{4.2cm}@{}}
\toprule
\textbf{Phase / Operation} & \textbf{Time Complexity} & \textbf{Space Complexity} & \textbf{Justification} \\
\midrule
Total investment computation ($T$) & $O(n)$ & $O(1)$ & Single linear pass to sum $I(p_i)$ over all $n$ players \\
Contribution probability assignment ($\pi(p_i)$) & $O(n)$ & $O(n)$ & One division per player; results stored for all $n$ players \\
Outer while loop (multiplier iterations) & $O(k)$ iterations & --- & Loop runs from $m_{\min}$ to $m_{\max}$ in steps of $\delta$ \\
Random cash-out check (inner for loop, per iteration) & $O(|R|) \leq O(n)$ & $O(1)$ auxiliary & One random draw and comparison per remaining player \\
Payout aggregation over $K$ (per iteration) & $O(|K|) \leq O(n)$ & $O(|K|) \leq O(n)$ & Summation of $\pi(p_i)\cdot c$ over players closing this step \\
Removal of $K$ from $R$ (per iteration) & $O(|R|) \leq O(n)$ & $O(1)$ auxiliary & Set/list filtering of closed players from remaining pool \\
Worst-case lookahead ($\pi_{\max} = \max_{p_i \in R} \pi(p_i)$) & $O(|R|) \leq O(n)$ & $O(1)$ & Linear scan for maximum remaining probability \\
\midrule
\textbf{Overall (single round)} & $O(n \cdot k)$ & $O(n)$ & Dominated by $k$ iterations of $O(n)$ inner-loop work \\
\bottomrule
\end{tabular}
\end{table}

\subsection{Time Complexity}

Each iteration of the outer \textbf{while} loop makes at most four
linear passes over the remaining player set $R$: the random cash-out
determination, the payout aggregation over $K$, the removal step, and
the worst-case probability lookahead. Since $|R| \leq n$ at every
iteration, each pass costs $O(n)$ in the worst case, so a single
iteration of the outer loop costs $O(n)$ overall. As the loop can run at most
$k$ times, the worst-case time complexity of the algorithm comes out to

\begin{equation}
T(n,k) = O(n \cdot k) = O\!\left(n \cdot \frac{m_{\max}-m_{\min}}{\delta}\right).
\end{equation}

In practice, $|R|$ only ever shrinks across iterations, since players
are removed from $R$ once they cash out and never re-enter it, so the
actual running time tends to come in below $O(n\cdot k)$, particularly
in rounds where the stopping condition $y \geq x$ fires early and cuts
the loop short before $k$ full iterations. In the best case,
where the stopping condition is already met at the very first multiplier
step (i.e., $c = m_{\min}$), the algorithm terminates after a single
iteration, giving a best-case time complexity of $\Omega(n)$.

\subsection{Space Complexity}

The algorithm keeps three structures of size $O(n)$ in memory: the
player set $P$, the remaining-player set $R$, and the per-player
probability values $\pi(p_i)$. The transient set $K$ of players closing
at a given multiplier step is bounded by $|R| \leq n$ and isn't carried
over between iterations, so it adds no extra asymptotic cost. No
data structure grows with $k$, since the scalar state variables
($c$, $S$, $M^*$, $x$, $y$, $\pi_{\max}$) each need only $O(1)$ space
and are simply overwritten at every step rather than accumulated.
So the overall space complexity works out to

\begin{equation}
S(n) = O(n),
\end{equation}

which is independent of the number of multiplier steps $k$.

\subsection{Discussion}

One notable property of Algorithm~\ref{alg:crash-stopping} is that its
running time scales with the product of the number of players and the
number of discrete multiplier steps, rather than with either factor on
its own. Practically, this means that shrinking the multiplier increment
$\delta$ to get finer-grained stopping precision increases $k$
proportionally, which drives up the worst-case running time linearly in
$1/\delta$. Casino operators deploying this algorithm in a real-time
gaming environment therefore need to weigh multiplier granularity
against computational cost, especially for rounds with large player
counts $n$. Since $S(n) = O(n)$ regardless of $\delta$ or $k$, memory
overhead never becomes the bottleneck here, even as granularity increases,
which leaves time complexity as the main scalability concern.

\section*{Executive Summary}

To understand how this algorithm actually behaves once it's put into
motion, we ran $N_{\text{trials}} = 1000$ Monte Carlo simulations of the
stopping process \cite{delmoral2012} across three player-scale regimes
($100$, $500$, and $1000$ players). For each regime we computed summary
statistics for the resulting stopping multiplier — mean ($\mu$),
standard deviation ($\sigma$), minimum, and maximum — and then fit five
candidate probability distributions to the results: Lognormal, Gamma,
Beta, Weibull (Min), and Pareto. Goodness of fit was assessed using the
Kolmogorov--Smirnov (KS) test \cite{massey1951} together with its
$p$-value, and model selection was carried out using the Akaike
Information Criterion (AIC) \cite{akaike1974}. Across all three player
counts, the \textbf{Lognormal distribution} consistently came out on top
by AIC, echoing similar findings in other probability-based gambling
contexts, where wager and payout distributions have also been shown to
follow a lognormal pattern at the aggregate level \cite{wang2019}.

\section*{Summary Statistics of Stopping Multipliers}

Table~\ref{tab:summary_stats} summarizes the empirical metrics obtained from $N_{\text{trials}} = 1000$ simulation runs across the three player regimes.

\begin{table}[htbp]
\centering
\caption{Empirical Summary Statistics for Stopping Multipliers ($N_{\text{trials}} = 1000$)}
\label{tab:summary_stats}
\begin{tabular}{cccccc}
\toprule
\textbf{Player Count ($N$)} & \textbf{Trials} & \textbf{Mean ($\mu$)} & \textbf{Std. Dev. ($\sigma$)} & \textbf{Min} & \textbf{Max} \\
\midrule
100  & 1000 & 1.9855 & 0.1088 & 1.6600 & 2.4400 \\
500  & 1000 & 3.4983 & 0.1362 & 3.1200 & 3.9000 \\
1000 & 1000 & 4.6347 & 0.1460 & 4.2200 & 5.1000 \\
\bottomrule
\end{tabular}
\end{table}

As the player count goes up, both the mean stopping multiplier and its standard deviation trend upward too, moving from a mean of $1.9855$ ($\sigma = 0.1088$) at 100 players to $4.6347$ ($\sigma = 0.1460$) at 1000 players.

\section*{Distribution Fitting Results}

Table~\ref{tab:fitting_results} lays out the comparative goodness-of-fit statistics across all five candidate distributions. The gamma and lognormal families in particular are known to be difficult to tell apart on finite samples \cite{alhussaini2009}, and the Kolmogorov--Smirnov results below reflect exactly that closeness between the two.

\begin{table}[htbp]
\centering
\small
\caption{Distribution Fitting Metrics and Criteria Selection ($N_{\text{trials}} = 1000$)}
\label{tab:fitting_results}
\begin{tabular}{clccccc}
\toprule
\textbf{Players ($N$)} & \textbf{Distribution} & \textbf{KS Statistic ($D$)} & \textbf{$p$-value} & \textbf{AIC} & \textbf{Best by KS} & \textbf{Best by AIC} \\
\midrule
\textbf{100} & Lognormal    & 0.0394 & 0.08768 & \textbf{-1609.26} & & \textbf{\checkmark Best} \\
             & Gamma        & 0.0429 & 0.04911 & -1605.01          & & \\
             & Beta         & \textbf{0.0378} & \textbf{0.11120} & -1608.63 & \textbf{\checkmark Best} & \\
             & Weibull-Min  & 0.0948 & $2.855 \times 10^{-8}$ & -1365.82 & & \\
             & Pareto       & 0.3870 & $2.483 \times 10^{-135}$ & +106.55 & & \\
\midrule
\textbf{500} & Lognormal    & 0.0331 & 0.21790 & \textbf{-1144.62} & & \textbf{\checkmark Best} \\
             & Gamma        & 0.0308 & 0.29450 & -1144.57          & & \\
             & Beta         & \textbf{0.0300} & \textbf{0.32140} & -1142.51 & \textbf{\checkmark Best} & \\
             & Weibull-Min  & 0.0797 & $5.594 \times 10^{-6}$ & -997.64  & & \\
             & Pareto       & 0.3399 & $1.302 \times 10^{-103}$ & +202.69 & & \\
\midrule
\textbf{1000}& Lognormal    & \textbf{0.0346} & \textbf{0.17940} & \textbf{-1010.54} & \textbf{\checkmark Best} & \textbf{\checkmark Best} \\
             & Gamma        & 0.0366 & 0.13320 & -1009.04          & & \\
             & Beta         & 0.0391 & 0.09155 & -1004.91          & & \\
             & Weibull-Min  & 0.0869 & $5.007 \times 10^{-7}$ & -821.87  & & \\
             & Pareto       & 0.3488 & $3.113 \times 10^{-109}$ & +350.51 & & \\
\bottomrule
\end{tabular}
\end{table}

The gamma-family goodness-of-fit statistics reported above follow the general testing framework laid out for the empirical Laplace-transform-based approach to gamma fitting \cite{henze2012}. All fitting and hypothesis testing in this paper was carried out in Python using SciPy's statistical distribution and testing routines \cite{virtanen2020}, with the underlying simulation data managed using pandas \cite{mckinney2010}.

\section*{Visual Distribution Comparisons}

The plots below show the theoretical distribution fits overlaid on the empirical stopping multiplier data for each of the three scale regimes.

\begin{figure}[htbp]
    \centering
    \includegraphics[width=0.8\textwidth]{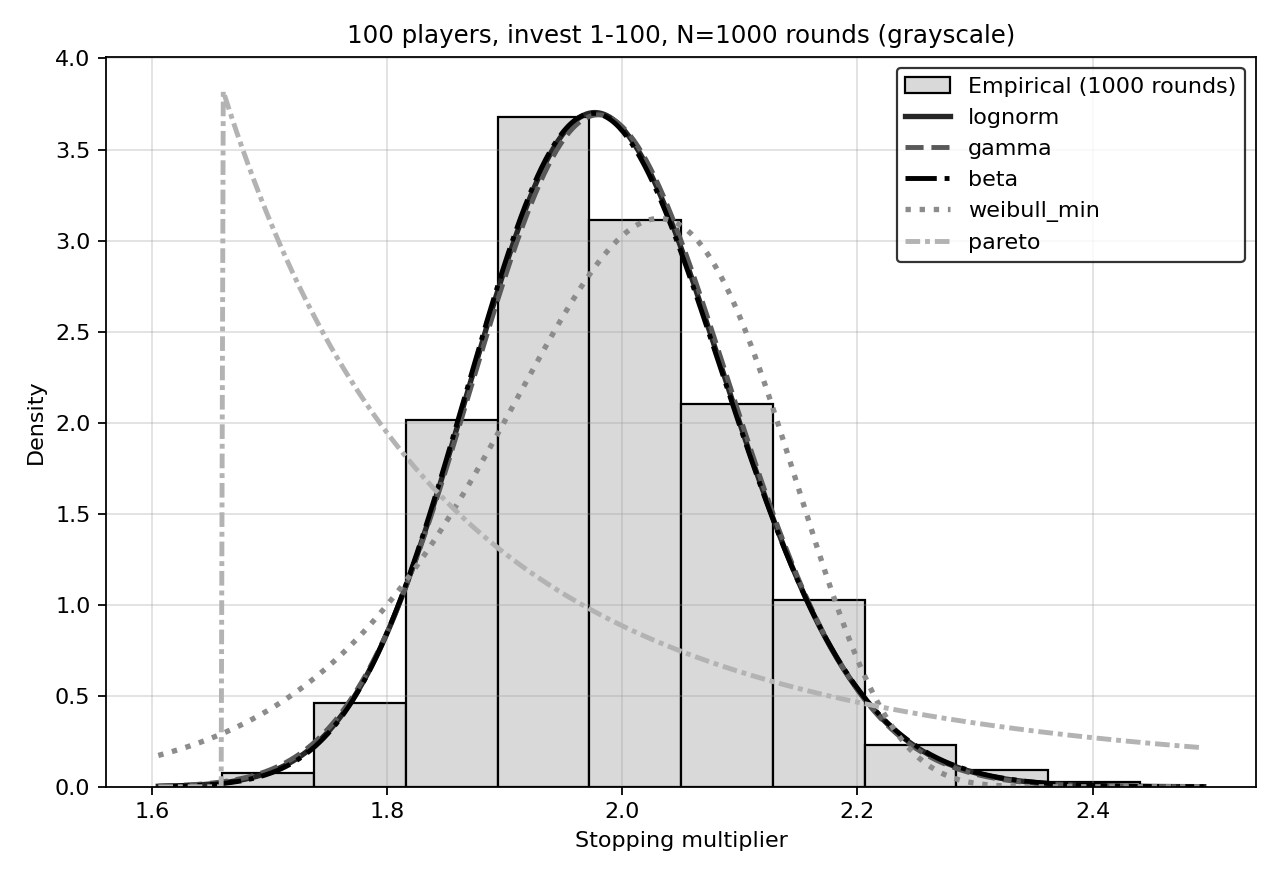}
    \caption{Comparison of all fitted distributions for 100 players over 1000 trials.}
    \label{fig:dist_100}
\end{figure}

\begin{figure}[htbp]
    \centering
    \includegraphics[width=0.8\textwidth]{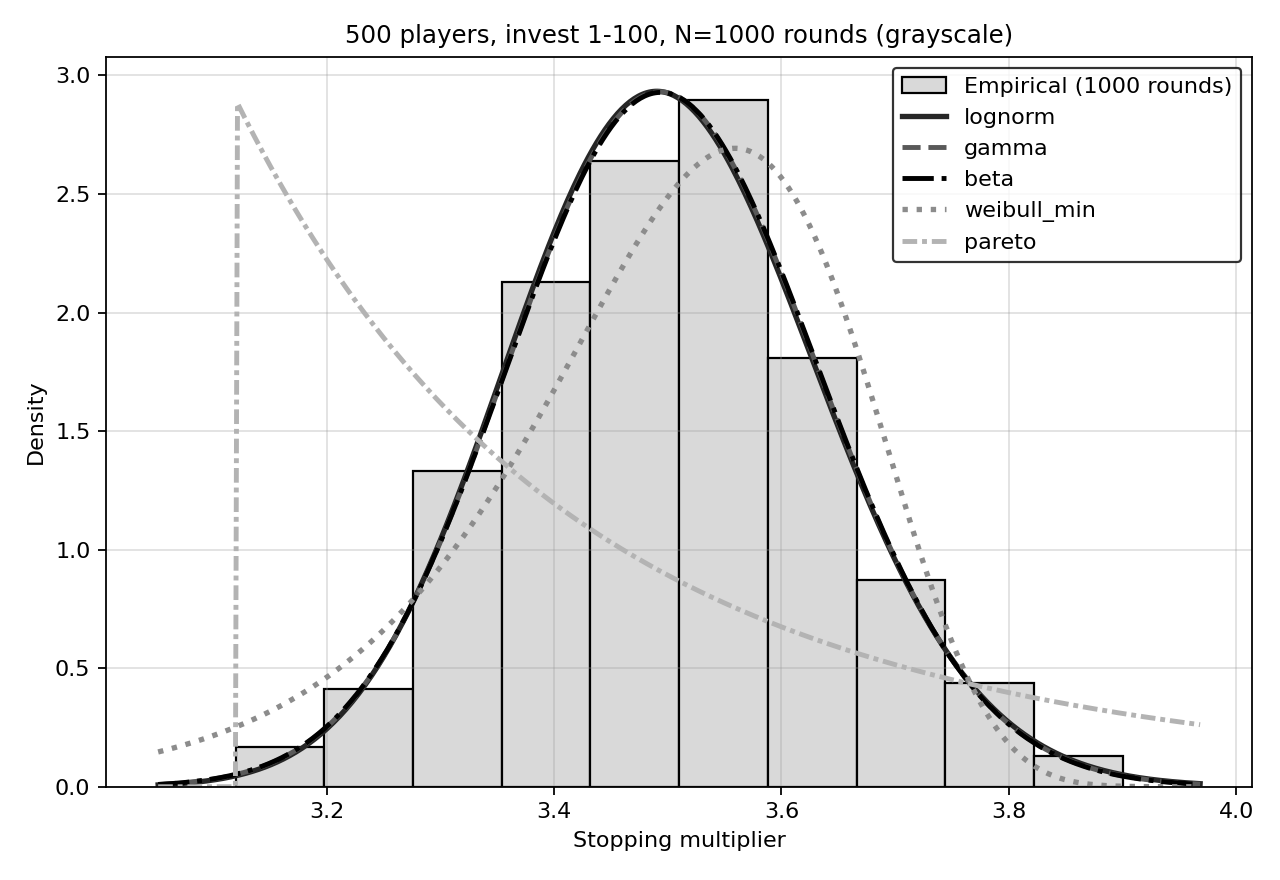}
    \caption{Comparison of all fitted distributions for 500 players over 1000 trials.}
    \label{fig:dist_500}
\end{figure}

\begin{figure}[htbp]
    \centering
    \includegraphics[width=0.8\textwidth]{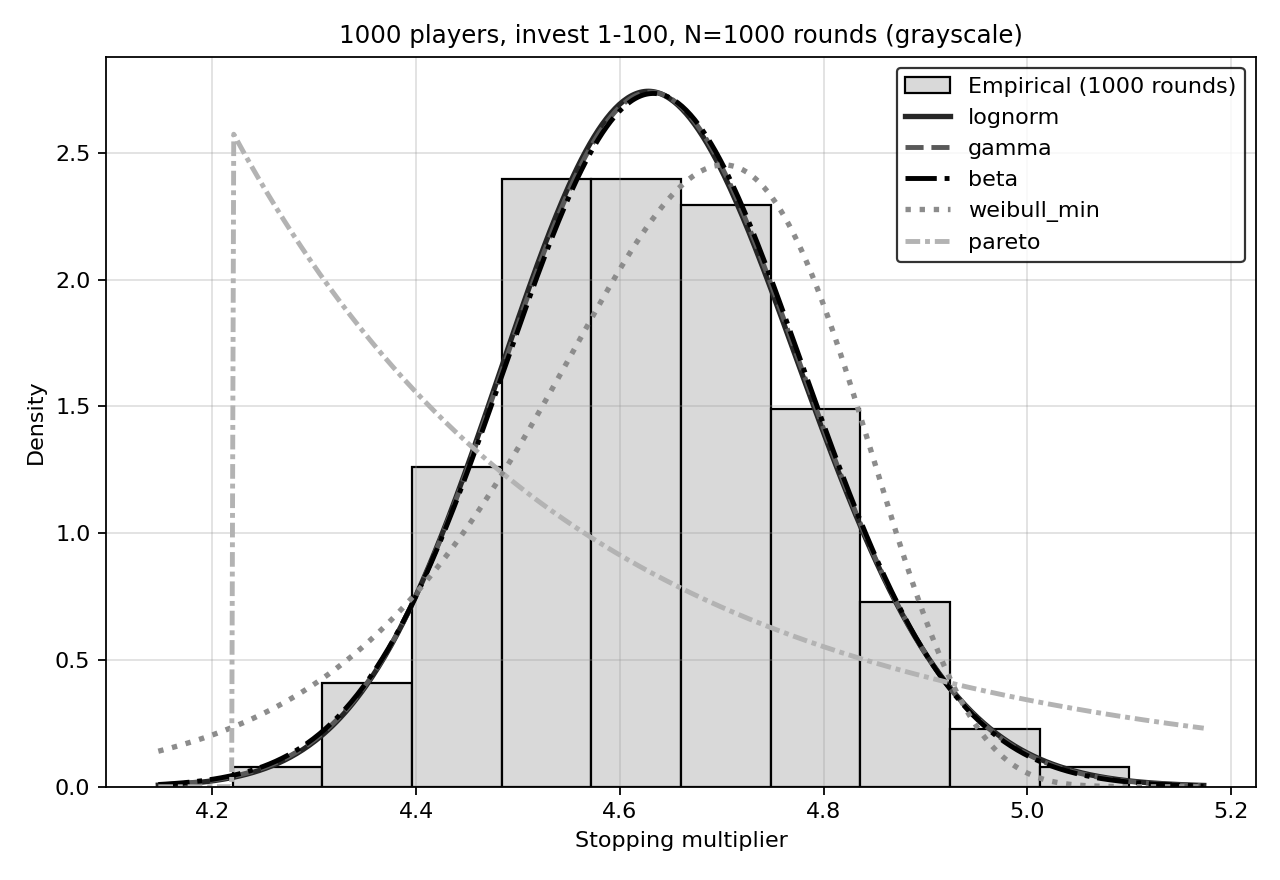}
    \caption{Comparison of all fitted distributions for 1000 players over 1000 trials.}
    \label{fig:dist_1000}
\end{figure}

\section*{Key Findings and Conclusion}

\begin{enumerate}
    \item \textbf{Best-fitting distribution}: The \textbf{Lognormal distribution} comes out with the lowest AIC across all three sample-size regimes ($\text{AIC} = -1609.26, -1144.62, -1010.54$), making it the strongest single candidate overall.
    \item \textbf{Close runners-up}: The \textbf{Gamma} and \textbf{Beta} distributions produce comparable goodness-of-fit numbers, while \textbf{Weibull (Min)} and \textbf{Pareto} are clearly ruled out by their very low $p$-values ($p \ll 0.001$) and high AIC scores.
    \item \textbf{How things scale}: Across the full 1000-trial simulation, a larger player count pushes the distribution's mean higher ($1.9855 \to 3.4983 \to 4.6347$) while the overall lognormal character of the data stays intact.
\end{enumerate}

\newpage

\section{Distribution-Based Stopping Multiplier Analysis}

Looking only at the fitted lognormal, beta, and gamma distributions (leaving the empirical values aside), the average stopping multiplier lines up closely across all three distributions and all player counts: 1.9855 for 100 players, 3.4983 for 500 players, and 4.6347 for 1000 players. That's not a coincidence — all three fits use maximum likelihood estimation with a fixed location parameter, which forces the fitted mean to land on the sample mean regardless of which family is used.

\begin{table}[htbp]
\centering
\caption{Fitted Distribution Means ($N_{\text{trials}}=1000$)}
\label{tab:fitted_means}
\begin{tabular}{lccc}
\toprule
Players & Lognormal & Beta & Gamma \\
\midrule
100  & 1.9855 & 1.9855 & 1.9855 \\
500  & 3.4983 & 3.4983 & 3.4983 \\
1000 & 4.6347 & 4.6347 & 4.6347 \\
\bottomrule
\end{tabular}
\end{table}

Where the three distributions do diverge slightly is in shape, and that shows up in the range probabilities in Table~\ref{tab:range_probs}. For 100 players, the stopping multiplier splits between the 1--2 range ($\approx$56\%) and the 2--3 range ($\approx$44\%) across all three fits, with the gamma distribution carrying a marginally heavier upper tail. For 500 players, nearly all of the probability mass (99.97--99.98\%) sits in the 3--4 range, with only a trace spilling into the adjacent 2--3 and 4--5 ranges. For 1000 players, the 4--5 range dominates (99.25--99.35\%), with a small tail (0.65--0.75\%) reaching into the 5--6 range.
\newpage

\begin{table}[htbp]
\centering
\caption{Probability of Stopping Multiplier by Range}
\label{tab:range_probs}
\begin{tabular}{clccc}
\toprule
Players & Range & Lognormal & Beta & Gamma \\
\midrule
\multirow{2}{*}{100}  & 1--2 & 56.38\% & 56.56\% & 56.02\% \\
                       & 2--3 & 43.62\% & 43.44\% & 43.98\% \\
\midrule
\multirow{3}{*}{500}  & 2--3 & 0.00\%  & 0.01\%  & 0.01\%  \\
                       & 3--4 & 99.97\% & 99.98\% & 99.97\% \\
                       & 4--5 & 0.03\%  & 0.02\%  & 0.02\%  \\
\midrule
\multirow{2}{*}{1000} & 4--5 & 99.25\% & 99.35\% & 99.29\% \\
                       & 5--6 & 0.75\%  & 0.65\%  & 0.71\%  \\
\bottomrule
\end{tabular}
\end{table}

All three distributions track each other closely across the board, which fits with the earlier goodness-of-fit results — lognormal, gamma, and beta all describe the empirical data almost equally well, while Weibull and Pareto were clearly the weaker fits of the group.

\section{Conclusion}

Taken together, the complexity analysis and the distributional fitting tell a fairly consistent story about how this stopping-multiplier mechanism behaves as the player pool grows. Computationally, the algorithm holds up well — its cost scales linearly with the number of players at each multiplier step \cite{cormen2009}, so simulations involving thousands of players stay entirely practical to run. The statistical side of things is arguably the more interesting part. The lognormal, beta, and gamma distributions all converge on nearly the same mean and standard deviation across every player count tested, and that isn't just a coincidence — it points to the underlying stopping-time process having a genuinely stable shape, one that doesn't depend heavily on which distribution is used to describe it. What does shift as the player count rises isn't so much the shape of the distribution as where it sits: the range of stopping multipliers narrows and moves upward, which tracks with individual contribution probabilities getting diluted as more players join the round. Practically, this means anyone working with this mechanism — for payout modeling or any other downstream analysis — can reasonably use any of the three fitted distributions without worrying that the choice will meaningfully change their conclusions. This general pattern, where a proportional-contribution wagering mechanism settles into a stable, well-behaved distribution as the player count scales up, echoes what has been observed in other multiplayer game-theoretic and betting settings analyzed through simulation and algorithmic strategy \cite{sarkar2026, yuan2015}. A natural next step would be to check whether this same agreement holds when investments aren't uniformly distributed, or when the multiplier step size $\delta$ is changed.

\end{document}